%% file: main.tex
\documentclass[
	a4paper, 
	10pt, 
	unnumberedsections, 
	twoside, 
]{LTJournalArticle}

\runninghead{} 

\footertext{\textit{RISC-V Summit Europe, Munich, 24-28th June 2024}} 

\title{Enabling an OpenStack-based cloud on top of RISC-V hardware.} 

\author{%
	Diego Marron\textsuperscript{1}
    \thanks{Corresponding author: \href{diego.marron@bsc.es}{\tt diego.marron@bsc.es}},  
    Aaron Call\textsuperscript{1}, Josep Ll. Berral\textsuperscript{2,1}, Ramon Nou\textsuperscript{1}
}

\date{\footnotesize\textsuperscript{\textbf{1}}Barcelona Supercomputing Center\\ \textsuperscript{\textbf{2}}Universitat Politècnica de Catalunya - BarcelonaTECH}

\renewcommand{\maketitlehookd}{%
	\begin{abstract}

The European Union's technological sovereignty strategy centers around the RISC-V Instruction Set Architecture, with the European Processor Initiative leading efforts to build production-ready processors. 
Focusing on realizing a functional RISC-V cloud ecosystem, the Vitamin-V European project developed an OpenStack cluster utilizing genuine hardware. 
In this poster, we detail the efforts done in porting and setting up the cluster and the many software services required by OpenStack to properly run on real hardware. 
In this poster, we detail our efforts on building an minimal viable prototype OpenStack cluster using real hardware. The cluster is almost functional, and we expect it to be complete in the next few months.
	\end{abstract}
}

\begin{document}

\maketitle 

\input{body}

\printbibliography 


\end{document}

%% file: body.tex
\section{Introduction and motivation}
\begin{figure*}[ht]
    \centering
    \includegraphics[scale=0.24]{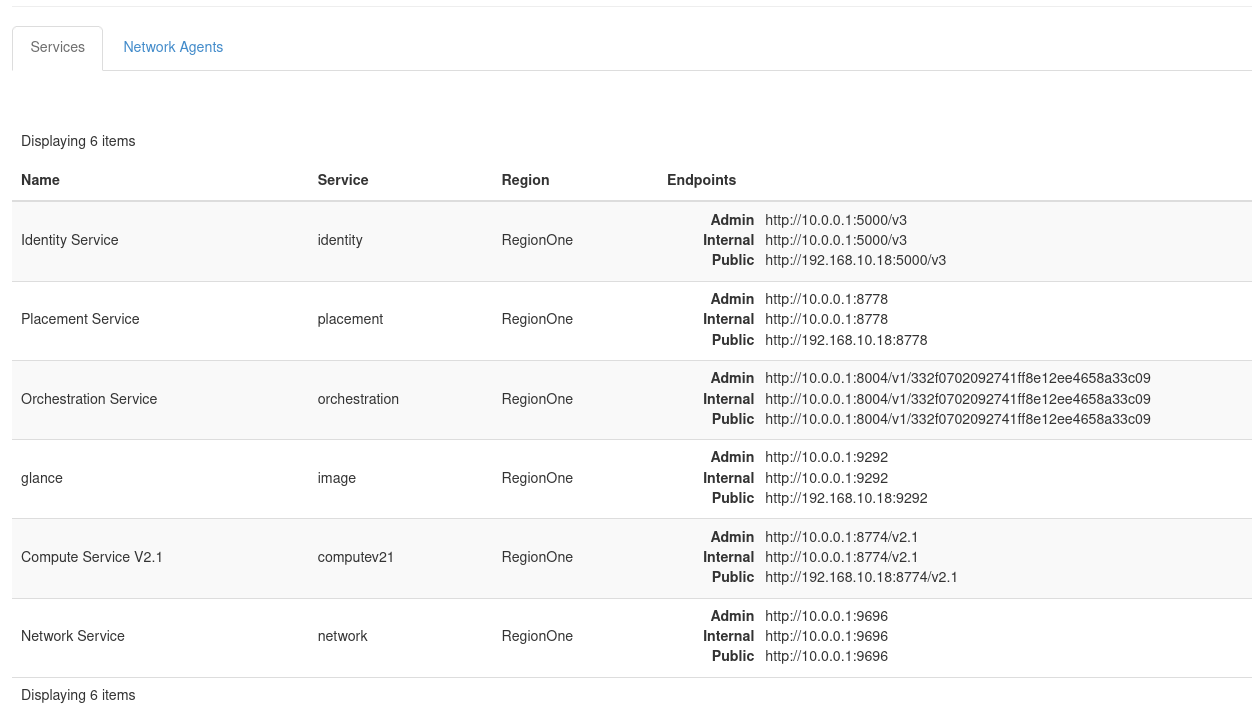}\\
    \caption{OpenStack Dashboard}
    \label{fig:os_dasbhoard}
\end{figure*}
The RISC-V Instruction Set Architecture (ISA) is at the core of the European Union's technological sovereignty plans. One key initiative in this direction is the European Processor Initiative (EPI)~\cite{epi}, 
which aims to produce processors ready for mass production based on the RISC-V open-source ISA. These processors are intended to be used in various applications, including cloud computing and data centers.

Large-scale supercomputers and datacenters become essential in many scientific applications requiring big-data processing: discovering relevant clinical, social, economic, or environmental indicators at any scale (e.g. 1 PetaByte of genomic datasets~\cite{omics_petabytes}). However, most current computing architectures are proprietary and closed-source technologies such as x86 and ARM, which creates concerns about the reliability of privacy and security. 

The Vitamin-V project emerged as an effort to build European data centers based on the EPI processor as a reliable alternative to traditional proprietary systems like x86 and ARM. As a part of the Vitamin-V project, we have devoted the last year to developing a functional OpenStack cluster utilizing real hardware instead of emulators. To this end, we are using a hardware platform based on RISC-V development boards. In this poster, we show how we are using Spieed Lichee PI 4A~\cite{licheepi} as a demonstrator in a RISC-V-based cloud environment.

In this poster, we explain our work to enable an OpenStack-based cloud environment on top of RISC-V using the Lichee PI 4A boards~\cite{licheepi}. Enabling such an environment allows us to demonstrate a functional cloud on RISC-V and evaluate how well traditional cloud workloads perform on top of it.

\subsection{Porting of OpenStack}
Due to the novelty of both RISC-V chips and OS (Linux), building the cluster from scratch presented several challenges at different levels:  operating system, distribution, and packages.

To run OpenStack, we built a custom Linux kernel to include missing modules needed by OpenStack (e.g., open switch) and removed unnecessary modules to make the custom kernel fit the partition. The default Linux distribution is a version of Debian by Sipeed that targets their repositories containing a small subset of all Debian packages (e.g., most development packages are missing). Regarding deployment and cluster management, we used Ansible also missing  in the Sipeed Debian repository.

As the project advanced, the official Debian repositories increased their package coverage up to 95\% of packages built for RISC-V, easing deployment procedures. We decided to switch all packages to official Debian, and only use the Sipeed software for building the custom kernel. 

Updating the operating system packages and configurations on all nodes is also challenging due to the maturity of the software. We use Ansible to manage the cluster, a popular tool among administrators for deploying and configuring large clusters to ensure reproducibility. We also explored using Devstack and Kolla to deploy and configure OpenStack. However, both options download specific versions of packages and dependencies, which turned into many compilation issues on RISC-V.

\begin{figure}[!b]
    \begin{center}
        \includegraphics[width=1\columnwidth]{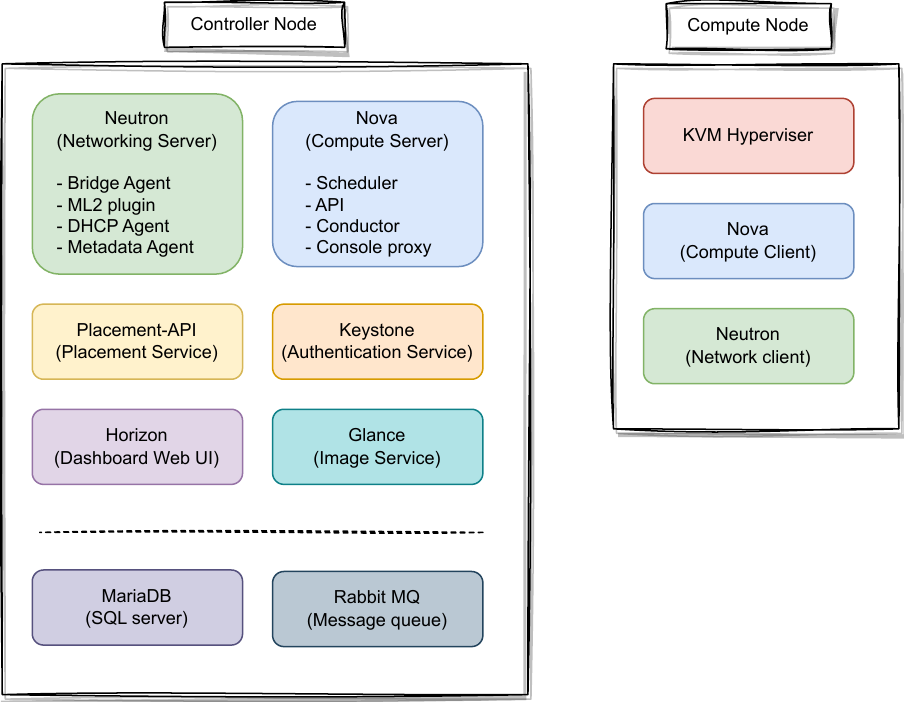}
        \caption{OpenStack software services}
        \label{fig:os_services}
    \end{center}
    \vspace{-3.5em}
\end{figure}

OpenStack consists of numerous software services, each exhibiting varying levels of maturity for the RISC-V architecture. While some components worked seamlessly on RISC-V, others necessitated an intricate setup process due to the absence of essential libraries. These often had to be constructed from source code. Figure \ref{fig:os_services} details various services running and their roles.

\subsection{Experimental platform}

\begin{figure}[!h]
    \begin{center}
        \includegraphics[width=0.55\columnwidth]{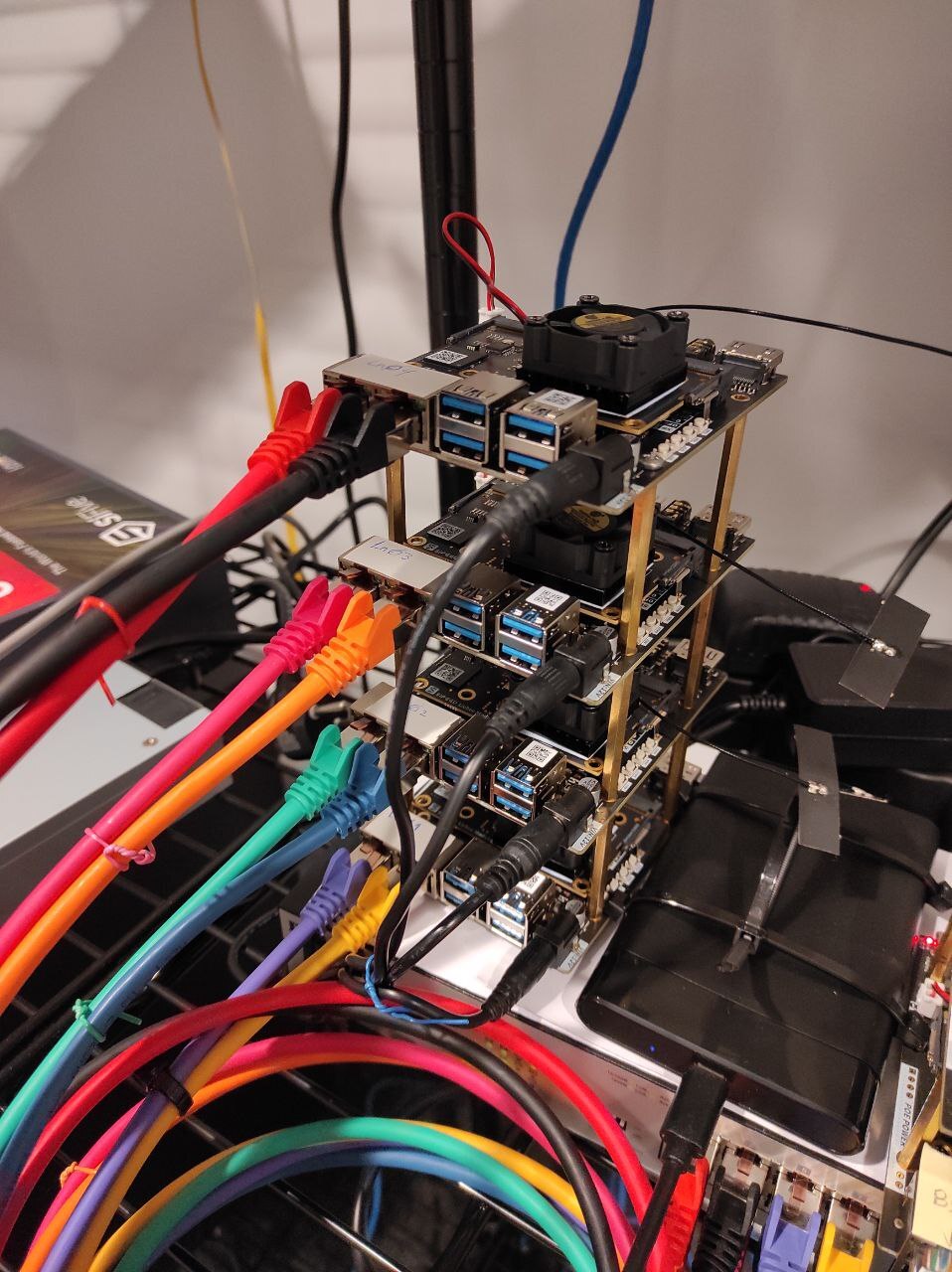}\
        \caption{Real-Hardware running OpenStack}
        \label{fig:os_licheenostrum}
    \end{center}
    \vspace{-2em}
\end{figure}

As the hardware platform, we are using a RISC-V development board by Sipeed, the Lichee PI 4A owing to their balanced pricing and capabilities. More specifically, the utilized development platform provides a TH1520 RISC-V CPU (4 Threads), 16GB of RAM, and 128GB of storage. In addition, it also provides a dual Gigabit Ethernet a feature particularly interesting when building an OpenStack cluster. Figure \ref{fig:os_licheenostrum} shows the cluster setup.

The current OpenStack cluster comprises 6 Lichee PI4A boards, each of which, takes the following roles: 1 controller node, 4 compute nodes, and 1 board serving files. The controller node runs the dashboard to launch new instances and set up machines. It also runs other services such as database, message queue, identity, and image services, along with the server side of Compute and Network. The compute and network services inform the hypervisor regarding the VM configuration. The configuration is set on the client side.

\subsubsection{Current status and next steps}
Currently, we have been able to run the basic OpenStack dashboard and we can identify the hardware resources available in the cluster. This dashboard is shown in figure \ref{fig:os_dasbhoard}. However, we are still troubleshooting issues regarding the virtual machine instantiation, which crashes. We do expect to successfully enable a functional OpenStack environment in the next few months and to start demonstrating its porting.

%% file: bibliography.bib
@String{Computing = "Computing" }

@article{omics_petabytes,
author={Eisenstein, Michael},
title={Big data: The power of petabytes},
journal={Nature},
year={2015},
month={Nov},
day={01},
volume={527},
number={7576},
pages={S2-S4},
issn={1476-4687},
doi={10.1038/527S2a},
url={https://doi.org/10.1038/527S2a}
}

@inproceedings{epi,
    author = {Kova\v{c}, Mario},
    title = {European Processor Initiative: The Industrial Cornerstone of EuroHPC for Exascale Era},
    year = {2019},
    isbn = {9781450366854},
    publisher = {Association for Computing Machinery},
    address = {New York, NY, USA},
    url = {https://doi.org/10.1145/3310273.3323432},
    doi = {10.1145/3310273.3323432},
    booktitle = {Proceedings of the 16th ACM International Conference on Computing Frontiers},
    pages = {319},
    numpages = {1},
    location = {Alghero, Italy},
    series = {CF '19}
}

@misc{licheepi,
    author = {SiPeed},
    howpublished = {\url{https://sipeed.com/licheepi4a}},
    note = {Accessed: 2024-01-25},
    title = {Sipeed Liche Pi4A}
}
